\documentstyle[epsfig]{aipproc}

\newcommand{\ra}{\rightarrow}

\begin{document}
\title{CP Violation in $\Lambda\rightarrow p \pi^-$: SM vs New Physics}

\author{G. Valencia}
\address{Department of Physics, Iowa State University, 
Ames, IA 50011\thanks{Supported in part by DOE under contract 
number DE-FG02-92ER40730.}}

\maketitle

\begin{abstract}

I discuss CP violation in $\Lambda \rightarrow p \pi^-$ 
comparing the standard model expectations with what could  
happen in new physics scenarios. I point out that Fermilab 
experiment E871 is sensitive to some of these scenarios.

\end{abstract}

\subsection*{Introduction}

In non-leptonic hyperon decays such as $\Lambda \rightarrow p \pi^-$ 
it is possible to search for CP violation by comparing the 
decay with the corresponding anti-hyperon decay \cite{op}. 
The Fermilab experiment E871 is currently searching for CP 
violation in such a decay and is sensitive to certain types 
of physics beyond the standard model. The observable provides 
information that is complementary to that obtained from the 
measurement of $\epsilon^\prime/\epsilon$.

The reaction of interest is the decay of a polarized $\Lambda$, 
with known polarization $\vec{w}$, into a proton (whose polarization is not 
measured) and a $\pi^-$ with momentum $q$. 
The final $p\pi^-$ state can be in an 
S-wave or a P-wave, and in an $I=1/2$ or $I=3/2$ state. The 
observables are the total decay rate and a correlation in the 
decay distribution of the form
\begin{equation}
{d\Gamma \over d\Omega}\sim 1 +\alpha \vec{w}\cdot\vec{q}
\end{equation}
The branching ratio for this mode is $63.9\%$ and the parameter 
$\alpha$ has been measured to be $\alpha = 0.64$ \cite{pdb}.
The CP violation in question involves a comparison of the 
parameter $\alpha$ with the corresponding parameter $\bar{\alpha}$ 
for the reaction $\bar{\Lambda}\ra \bar{p}\pi^+$.

It is standard to write the amplitudes in terms of their isospin 
components in the form
\begin{eqnarray}
S &=& S_1 e^{i\delta_1^S} + S_3 e^{i\delta_3^S} \nonumber \\
P &=& P_1 e^{i\delta_1^P} + P_3 e^{i\delta_3^P} 
\label{ampdef}
\end{eqnarray}
A $\Delta I =1/2$ rule is observed experimentally, 
$S_3/S_1 \approx 0.026$ and $P_3/P_1 = 0.03\pm 0.03$ \cite{over}. 
The strong $\pi N$ scattering phases have been measured for the 
$I=1/2$ channel, $\delta^S_1 \sim 6^o$ and $\delta^P_1 \sim -1^o$ 
\cite{roper}. 
The $I=3/2$ scattering phases have been measured with large errors 
but are not needed here.

To discuss CP violation, we allow the amplitudes in Eq.~\ref{ampdef}
to have a CP violating weak phase, $S_i \ra S_i \exp(i\phi^S_i)$ 
and $P_i \ra P_i \exp(i\phi^P_i)$ and compare the pair of 
CP conjugate reactions. CP symmetry predicts that $\Gamma = 
\bar{\Gamma}$ and that $\bar{\alpha}=-\alpha$. One therefore 
defines the CP-odd observables
\begin{eqnarray}
\Delta &\equiv &{\Gamma -\bar\Gamma \over \Gamma + \bar\Gamma}
\sim  \sqrt{2}{S_3\over S_1}\sin(\delta^S_3-\delta^S_1)
\sin(\phi^S_3-\phi^S_1) \nonumber \\
A(\Lambda^0_-) &\equiv & {\alpha + \bar\alpha\over \alpha-
\bar\alpha}
\sim  -\sin(\delta^P_1-\delta^S_1)
\sin(\phi^P_1-\phi^S_1) 
\sim  0.12 \sin(\phi^P_1-\phi^S_1)
\label{asym}
\end{eqnarray} 
The partial rate asymmetry is very small, being suppressed by 
three small factors, $S_3/S_1$, strong phases, and weak phases. 
It represents an interference between amplitudes with $\Delta I =1/2$ 
and $\Delta I =3/2$. The asymmetry $A(\Lambda^0_-)$, on the other hand, 
is not suppressed by the $\Delta I =1/2$ rule, as it originates in an 
interference of S and P-waves within the $\Delta I =1/2$ transition. 
For this reason, the observable $A(\Lambda^0_-)$ is {\it qualitatively} 
different from $\epsilon^\prime/\epsilon$.

The experiment E871 at Fermilab produces the polarized $\Lambda$ 
from the weak decay $\Xi^-\ra \Lambda \pi^-$ and for this reason 
what they measure is actually the combination 
$A(\Lambda^0_-)+A(\Xi^-_-)$. Their expected sensitivity is $10^{-4}$. 
The weak phases in $\Xi^-$ decay 
(within the standard model) have been estimated to be about 
two times smaller than those in $\Lambda$ decay \cite{dhp}. Similarly, the 
strong phases in $\Xi^-$ decay are estimated to be of order 
$1^o$ \cite{lsw,dp} and therefore five times smaller than the 
strong phase difference in $\Lambda$ decay. For these two reasons 
we expect that the E871 measurement will be dominated by 
$A(\Lambda^0_-)$.

\subsection*{Standard Model}

Within the standard model one writes the $|\Delta S|=1$ effective 
weak Hamiltonian as a sum of four-quark operators multiplied by 
Wilson coefficients in the usual way,
\begin{equation}
H={G_F\over \sqrt{2}}V^*_{ud}V_{us}\sum_{i=1}^{12}c_i(\mu)
Q_i(\mu)
\end{equation}
This is, of course, the same effective Hamiltonian responsible for 
Kaon non-leptonic decays and is very well known. In particular the 
Wilson coefficients, $c_i(\mu)$ have been calculated in detail 
by Buras and his collaborators \cite{bbh}. The remaining problem is to 
calculate the matrix elements of the four-quark operators between 
hadronic states. This problem has not been resolved yet, and there 
is large theoretical uncertainty in these matrix elements. The 
usual way to proceed (which is the same as in kaon physics) is to 
take the real part of the matrix element from experiment 
(assuming CP conservation) and to use the calculated imaginary parts. 

Unlike the case of $\epsilon^\prime$, where both $\Delta I =1/2,3/2$ 
amplitudes are important, $A(\Lambda^0_-)$ is dominated by CP 
violation in $\Delta I =1/2$ amplitudes. One expects that the 
asymmetry will be dominated by the penguin operator with small 
corrections from other operators. A detailed study using vacuum 
saturation to estimate the matrix elements supports the view that 
$Q_6$ is dominantly responsible for $A(\Lambda^0_-)$ \cite{hsv}.

\begin{figure}[hctb]
\centerline{\psfig{figure=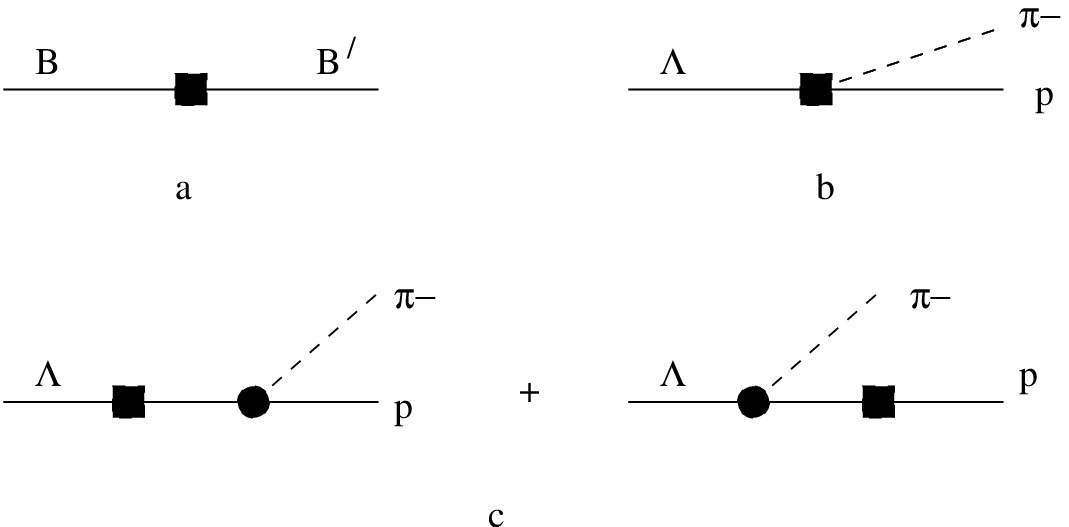}}
\caption{a) $B\ra B^\prime$ transition due to $Q_6$, solid square. 
b) S-wave obtained from (a) via a soft-pion theorem. c) P-wave 
obtained from (a) with strong pion emission (solid circle).}
\label{fig1}
\end{figure}

Once we have determined that only $Q_6$ is important, the 
strategy is to calculate the matrix elements of the form 
$<B'|Q_6|B>$ using a model, and then use these results to 
treat the non-leptonic hyperon decay at leading order in chiral 
perturbation theory as sketched in Figure~1. 
Equivalently, the S-waves are obtained 
with a soft-pion theorem and the P-waves with baryon poles. 
At present, the baryon to baryon matrix elements are taken from 
the MIT bag model calculation of Ref.~\cite{dghp}. 

It is difficult to quantify the theoretical error in this 
calculation. There are the obvious uncertainties in the 
short distance parameters as well as errors in the value of 
the strong phases. However, of greater 
concern is the issue of assigning an error to the hadronic matrix 
elements. Even if we assume that the baryon to baryon matrix elements 
calculated in the MIT bag model are exact, we know from the study 
of CP conserving amplitudes that non-leading order terms in 
chiral perturbation theory can be as large as the leading order 
amplitudes. For example, the s-wave imaginary part calculated in vacuum 
saturation, is a higher order correction to the bag-model plus soft 
pion theorem amplitude outlined above, but it is larger \cite{hsv}. To get an 
idea for the impact of this error we assign 
an overall error of a factor of two to the calculated matrix elements 
plus an overall 30\% uncorrelated error between S and P-waves. 
Combining all this results in, 
\begin{equation}
A(\Lambda^0_-) = (-3.0 \pm 2.6)\times 10^{-5}.
\end{equation}

\subsection*{Beyond the Standard Model}

There have been several estimates of $A(\Lambda^0_-)$ 
beyond the standard model. For the most part these 
studies discuss specific models, concentrating on one or 
a few operators and normalizing the strength of CP violation 
by fitting $\epsilon$. Some of these results (which have not 
been updated to incorporate current constraints on model 
parameters) are:
\begin{equation}
A(\Lambda^0_-) = \cases{-2\times 10^{-5} & SM \cite{dhp} \cr
                        -2\times 10^{-5} & 3 Higgs \cite{dhp} \cr 
                         0               & Superweak \cr
                         6\times 10^{-4} & LR \cite{chp}\cr}
\end{equation}

Perhaps a more interesting question is whether it is 
possible to have large CP violation in hyperon decays 
in view of what is known about $\epsilon$ and $\epsilon^\prime$. 
This question has been addressed in a model independent way 
by considering all the CP violating operators that can be 
constructed at dimension 6 that are compatible with the 
symmetries of the standard model \cite{hvop}. 
With this general formalism 
one can compute the contributions of each new 
CP violating phase to $\epsilon, \epsilon^\prime$,~and~$A(\Lambda^0_-)$. 
Of course, there is the caveat that the hadronic matrix 
elements cannot be computed reliably. Nevertheless, one 
finds in general, that parity even operators generate a 
weak phase $\phi_1^P$ and do not contribute to $\epsilon^\prime$. 
Their strength can be bound from the long distance 
contributions to $\epsilon$ that they induce. Similarly, the 
parity-odd operators generate a weak phase $\phi_1^S$ and 
contribute to $\epsilon^\prime$ (but not to $\epsilon$). 

The constraints from $\epsilon^\prime$ turn out to be much more stringent 
than those from $\epsilon$, and, therefore, the only 
natural way (without invoking fine cancellations between 
different operators) to obtain a large $A(\Lambda^0_-)$ 
given what we know about $\epsilon^\prime$ is with new 
CP-odd, P-even interactions. Within the model independent 
analysis, one can identify a few new operators with the 
required properties, that can lead to \cite{hvop},
\begin{equation}
A(\Lambda^0_-) \sim 5 \times 10^{-4} ~~~{\rm P-even, CP-odd}
\end{equation}

This possibility has been revisited recently, motivated in 
part by the observation of $\epsilon^\prime$. The 
average value $\epsilon^\prime/\epsilon = (21.2 \pm 4.6)\times 10^{-4}$ 
\cite{epp} appears 
to be larger than the standard model central prediction with simplistic 
models for the hadronic matrix elements. This has motivated 
searches for new sources of CP violation that can give 
large contributions to $\epsilon^\prime$, in particular, 
within supersymmetric theories. One such scenario generates 
a large $\epsilon^\prime$ through an enhanced gluonic 
dipole operator \cite{mm}. The effective Hamiltonian is of the form
\begin{eqnarray}
H_{eff}&=&(\delta_{12}^d)_{LR}C_g\bar{d}\sigma_{\mu\nu}t^a
(1+\gamma_5 )sG^{a\mu\nu}\nonumber \\
&+&
(\delta_{12}^d)_{RL}C_g\bar{d}\sigma_{\mu\nu}t^a
(1-\gamma_5 )sG^{a\mu\nu}
\end{eqnarray}
The quantity $C_g$ is a known loop factor, and the 
$(\delta_{12}^d)_{LR,RL}$ originate in the supersymmetric 
theory \cite{ggms}. Depending on the correlation between the 
value of $(\delta_{12}^d)_{LR}$ and $(\delta_{12}^d)_{RL}$ 
one gets different scenarios for $\epsilon^\prime$ and 
$A(\Lambda^0_-)$ as shown in Figure~2 \cite{hmpv}.
\begin{figure}[hctb]
\centerline{\psfig{figure=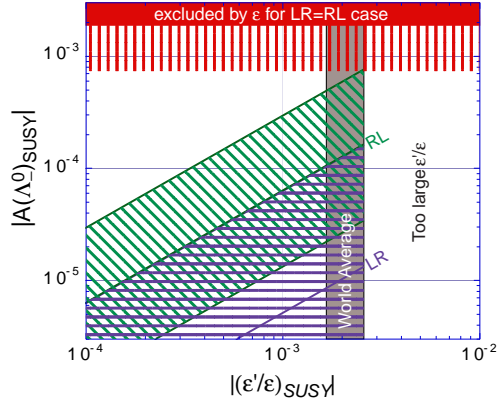,width=6.7cm}}
  \caption{The allowed regions on $(|(\epsilon'/\epsilon)_{SUSY}|,$ 
    $|A(\Lambda^{0}_-)_{SUSY}|)$ parameter space for three cases: a)
    only ${\rm Im}(\delta^d_{12})_{LR}$ contribution, which is the
    conservative case (hatched horizontally), b) only ${\rm
      Im}(\delta^d_{12})_{RL}$ contribution (hatched diagonally), and
    c) ${\rm Im}(\delta^d_{12})_{LR}= {\rm Im}(\delta^d_{12})_{RL}$
    case which does not contribute to $\epsilon'$ and can give a large
    $|A(\Lambda^{0}_{-})|$ below the shaded region (or vertically
    hatched region for the central values of the matrix elements).  The
    last case is motivated by the relation $\lambda =
    \sqrt{m_{d}/m_{s}}$.  The vertical shaded band is the world
    average \protect\cite{epp} of $\epsilon'/\epsilon$.  The region to
    the right of the band is therefore not allowed.}
\label{fig2}
\end{figure}
For example, if only $(\delta_{12}^d)_{LR}$ 
is non-zero, there can be a large $\epsilon^\prime$ \cite{mm}, 
but $A(\Lambda^0_-)$ is small as in the 3-Higgs model of 
\cite{dhp}. However, in models in which 
${\rm Im}(\delta_{12}^d)_{LR}={\rm Im}(\delta_{12}^d)_{RL}$ 
the CP violating operator is parity-even. In this case there 
is no contribution to $\epsilon^\prime$ and $A(\Lambda^0_-)$ 
can be as large as $10^{-3}$ \cite{hmpv}. It is interesting that 
this type of model is not an ad-hoc model to give a large 
$A(\Lambda^0_-)$, but is a type of model originally designed 
to naturally reproduce the relation $\lambda =\sqrt{m_d/m_s}$, 
as in Ref.~\cite{bdh}, for example.

\subsection*{Conclusion and Comments}

E871 is expected to reach a sensitivity of $10^{-4}$ for the 
observable $A(\Lambda^0_-)+A(\Xi^-_-)$.

\begin{itemize}

\item $A(\Lambda^0_-)$ is likely to be significantly 
larger than $A(\Xi^-_-)$.

\item $A(\Lambda^0_-) = (-3.0 \pm 2.6)\times 10^{-5}$ is our 
current best guess for the standard model and the theoretical 
uncertainty is dominated by our inability to calculate hadronic 
matrix elements reliably. For this reason, the error assigned to this 
quantity is no more than an educated guess.

\item $A(\Lambda^0_-)$ can be much larger if CP violation 
originates in P-even new physics. A specific realization of this 
scenario is possible in supersymmetric theories leading 
to $A(\Lambda^0_-)$ as large as $10^{-3}$.

\end{itemize}

I conclude that a non-zero measurement by E871 is not only 
possible but that it would provide valuable complementary 
information to what we already know from $\epsilon^\prime$. 

Finally I would like to mention two related issues. 
A search for $\Delta S =2$ 
hyperon non-leptonic decays is also a useful enterprise as 
it provides information that is complementary to what we 
know from $K-\bar{K}$ mixing \cite{hvds}. A CP violating rate asymmetry 
in $\Omega \rightarrow \Xi \pi$ decay can be as large as 
$2\times 10^{-5}$ within the standard model (and up to ten times larger 
beyond), much larger than the corresponding rate asymmetries in 
octet-hyperon decay \cite{tv}.

This work was supported by DOE under contract number 
DE-FG02-92ER40730. This talk summarizes work done in  
collaboration with  John Donoghue, Xiao-Gang He, Hitoshi Murayama, 
Sandip Pakvasa, Herbert Steger and Jusak Tandean.


\begin{thebibliography}{9}

\bibitem{op} S. Okubo, Phys. Rev. {\bf 109}, 984 (1958); 
A. Pais, Phys. Rev. Lett. {\bf 3}, 242 (1959).

\bibitem{pdb} C.~Caso {\it et al.},
{\it Eur.\ Phys.\ J.\ }  {\bf C3}, 1 (1998).

\bibitem{over} O.~E.~Overseth, in Review of Particle Properties, 
{\it Phys. Lett.} {\bf 111B}, 286 (1982). 

\bibitem{roper} L. D. Roper, R. M. Wright and B. Feld, 
Phys. Rev. {\bf 138}, 190 (1965); A. Datta and S. Pakvasa, Phys. Rev. 
{\bf D56}, 4322 (1997).

\bibitem{dhp} J. Donoghue and S. Pakvasa, Phys. Rev. Lett. 
{\bf 55}, 162 (1985);
J. Donoghue, X.-G. He and S. Pakvasa, Phys. Rev. {\bf D34}, 833 (1986).

\bibitem{lsw} M. Lu, M. Savage and M. Wise, Phys. Lett. {\bf B337}, 
133 (1994).

\bibitem{dp} A. Datta and S. Pakvasa, Phys. Lett. {\bf B344}, 430 (1995); 
A. Kamal, Phys. Rev. {\bf D58}, 077501 (1998). 

\bibitem{bbh} G. Buchalla, A. Buras and M. Harlander, Nucl. Phys. 
{\bf B337}, 313 (1990); G.~Buchalla, A.~J.~Buras and M.~E.~Lautenbacher,
{\it Rev.\ Mod.\ Phys.\ } {\bf 68}, 1125 (1996).

\bibitem{hsv} X.-G. He, H. Steger, and G. Valencia, Phys. Lett. {\bf B272},  
411 (1991).

\bibitem{dghp} J. Donoghue {\it et. al.}\/, Phys. Rev. {\bf D23}, 1213 
(1981).

\bibitem{chp} D. Chang, X.-G. He and S. Pakvasa, Phys. Rev. Lett. 
{\bf 74}, 3927 (1995).

\bibitem{hvop} X.-G. He and G. Valencia, Phys. Rev. {\bf D52}, 5257 (1995).  

\bibitem{epp} This is the average of E731, NA31, KTeV and NA48 with the
  error bar inflated to obtain $\chi^2/{\rm d.o.f.} = 1$ according to
  the Particle Data Group prescription.

\bibitem{mm} A.~Masiero and H.~Murayama, Phys. Rev. Lett. {\bf 83}, 907 
(1999).

\bibitem{ggms} F. Gabbiani, {\it et. al.}, Nucl. Phys. {\bf B447}, 
321 (1996).

\bibitem{hmpv} Xiao-Gang~He, H.~Murayama, S.~Pakvasa and G.~Valencia,
{\it Phys.\ Rev.\ } {\bf D61}, 071701 (2000).

\bibitem{bdh} R. Barbieri, G. Dvali and L.J. Hall, Phys. Lett. {\bf 
B377}, 76 (1996).

\bibitem{hvds} X.-G. He and G. Valencia, Phys. Lett. {\bf B409}, 
469 (1997) Erratum-ibid. {\bf B418}, 443 (1998). 

\bibitem{tv} J. Tandean and G. Valencia, 
Phys. Lett. {\bf B451}, 382 (1999). 


\end{thebibliography}
\end{document}